\documentclass[11pt,preprint]{aastex}
\usepackage{graphics}
\usepackage{psfig}
\usepackage{longtable}
\usepackage{psfig}
\def\gsim{\lower 2pt \hbox{$\, \buildrel {\scriptstyle >}\over
{\scriptstyle \sim}\,$}}
\def\lsim{\lower 2pt \hbox{$\, \buildrel {\scriptstyle <}\over
{\scriptstyle \sim}\,$}}

\begin{document}
\def\xs{{Abell 2125}}
\title{Correction for the Flux Measurement Bias in X-ray Source Detection}
\author{Q. Daniel Wang}
\affil{Astronomy Department, University of Massachusetts, Amherst, MA 01003,USA}\affil{Email: wqd@astro.umass.edu}
\shortauthors{Wang}
\shorttitle{}
\begin{abstract}
With a high spatial resolution imaging instrument such as the 
{\sl Chandra}/ACIS, one can confidently identify 
an X-ray source with only a few detected counts. 
The detection threshold of such sources, however, varies
strongly across the field-of-view of the instrument. 
Furthermore, the low detection counting statistics, 
together with a typical steep source number-flux relation, 
causes more intrinsically faint
sources to be detected at apparently higher fluxes than the other
way around. We quantify this ``X-ray Eddington bias'' 
as well as the detection threshold variation and devise 
simple procedures for their corrections. 
To illustrate our technique, we present results from our analysis of 
X-ray sources detected in the fields of the large-scale 
hierarchical complex Abell 2125 at $z = 0.247$ and the nearby galaxy NGC 4594
(Sombrero). We show that the sources detected in the Abell 2125 
field, excluding 10 known complex members, have a number-flux relation 
consistent with the expected from foreground or background objects. 
In contrast, the 
number-flux relation of the NGC 4594 field is dominated by X-ray sources
associated with the galaxy. This galactic  component  of the relation
is well characterized by a broken power law.

\end{abstract}


\section{Introduction}
A classic problem in studying faint sources is the 
determination of both the threshold of their detection and the bias 
in their flux measurement. This problem is particularly acute
in the study of discrete X-ray sources, most of which 
are detected with very limited counting statistics in a typical
observation. Furthermore, the threshold may vary strongly 
across the field of view (FoV) of the photon-detecting instrument, depending on
the point spread function (PSF; e.g., Jerius et al. 2000) 
as well as the local background and 
effective exposure. The threshold variation also differs from one
observation to another but is often overlooked or oversimplified.

Closely related to the detection threshold
is the bias in the measurement of source fluxes. 
Because of the statistical uncertainties in the photon counting,
the fluxes are statistically over-estimated because there are 
typically far more truly faint sources than bright ones. 
So more sources are ``up-scattered'' to a given flux measurement 
than those that are ``down-scattered'', which is similar to the 
so-called Eddington bias
in the optical photometry of faint objects (Eddington 1940; 
Murdoch, Crawford, \& Jauncey 1973; Hogg \& Turner 1998;
 Kenter \& Murray 2003 and references therein).
Hogg \& Turner (1998) prescribed a maximum likelihood (ML) correction 
for the bias as a function of the detection 
signal-to-noise ratio  (S/N), assuming a Gaussian-distributed 
error in the photometry. They further concluded that a source identified 
with S/N $\sim 4$ or less is
practically useless, because the bias would be too 
severe to allow for any reasonable estimation of the true flux.
In X-ray astronomy, source detection thresholds corresponding to 
S/N $\sim 3$ or less are often used (e.g., Di Stefano
et al. 2003). But because the counting error distribution 
here is typically Poissonian, rather than Gaussian, a S/N ratio 
alone does not directly translate to a false detection probability, 
which also depends on the local background contribution 
(Schmitt \& Maccacaro 1986). Therefore, it
is important to check how this difference in the error distributions 
affects the Eddington bias. 

One can effectively treat both the threshold variation and the Eddington bias by calculating a redistribution matrix 
of the source flux measurement (Kenter \& Murray 2003). The analysis of the
source number-flux relation, for example, is mathematically analogous to 
spectral fitting, for which sophisticated software packages such as 
XSPEC  (Arnaud 1996) are publicly available and are widely used. 
With such a software package, one can effectively 
analyze the number-flux relation of X-ray sources, including the
effects of both the threshold variation and the Eddington bias. 
Indeed, Kenter \& Murray (2003) have recently illustrated this technique by 
using extensive ray-tracing and Monte Carlo simulations and by applying it to the
wavelet source-detection analysis of a {\sl Chandra} ACIS image. 
Here we describe a simple and yet general procedure that allows for the calculation of 
the  redistribution matrix without resorting to the ray-tracing.

To illustrate this procedure, we 
present the analysis of X-ray sources 
detected in two {\it Chandra} observations. The first is an 
observation of Abell 2125 (Fig. 1), a complex of 
galaxies and diffuse hot gas 
at $z = 0.247$. This 82 ksec observation
 has the FoV of 17$^\prime \times$ 17$^\prime$ (including
only the 2$\times$2 ACIS-I CCD array). Wang, Owen, \& Ledlow (2004) have presented
the main results from the observation, which are based partly on
the analysis discussed here. The second observation is 
an 18.5 ksec ACIS-S exposure of NGC 4594 (Sombrero), a nearly 
edge-on Sa galaxy at a distance of 8.9 Mpc (Fig. 2). 
Only the data from the on-axis (\# 7)  back-illuminated chip with a FoV of 
8\farcm4$\times$8\farcm4 
are included. A study of the discrete X-ray sources detected
with the same data has been reported by Di Stefano et al. (2003) and 
is compared in the present work. 

Our study does not account for all potential instrumental effects.
In particular, multiple sources of small angular separations
may produce a single detection, affecting not 
only the source number counting but also the shape of the number-flux relation
(Hasinger et al. 1993). Fortunately, with the superb spatial resolution of
an imaging instrument such as {\sl Chandra}, the effect is typically not 
important ($\lesssim $ a few percent). For the observations analyzed  here, 
we use the 
position-dependent 90\% energy-encircled radius  as the detection
aperture (EER; Jerius et al. 2000), which is a factor of 2 smaller than the 
source removal radius shown in Figs. 1 and 2b. Clearly, few sources are
affected by overlapping detection apertures.

\section{Brief Discussion of X-ray Source Detection and Measurement}

Here we consider the detection of point-like 
X-ray sources based on an event count image (e.g., Figs. 1 and 2b).
Counts in such an image can be divided approximately into two 
components: one from individual X-ray sources and the other from a 
smoothly-distributed background, consisting
of X-ray photons and charged particle-induced events. This background
varies across the image, but typically on scales substantially greater
than the size of the PSF (Jerius et al. 2000). 

\begin{figure}[htb!]
\centerline{
\psfig{figure=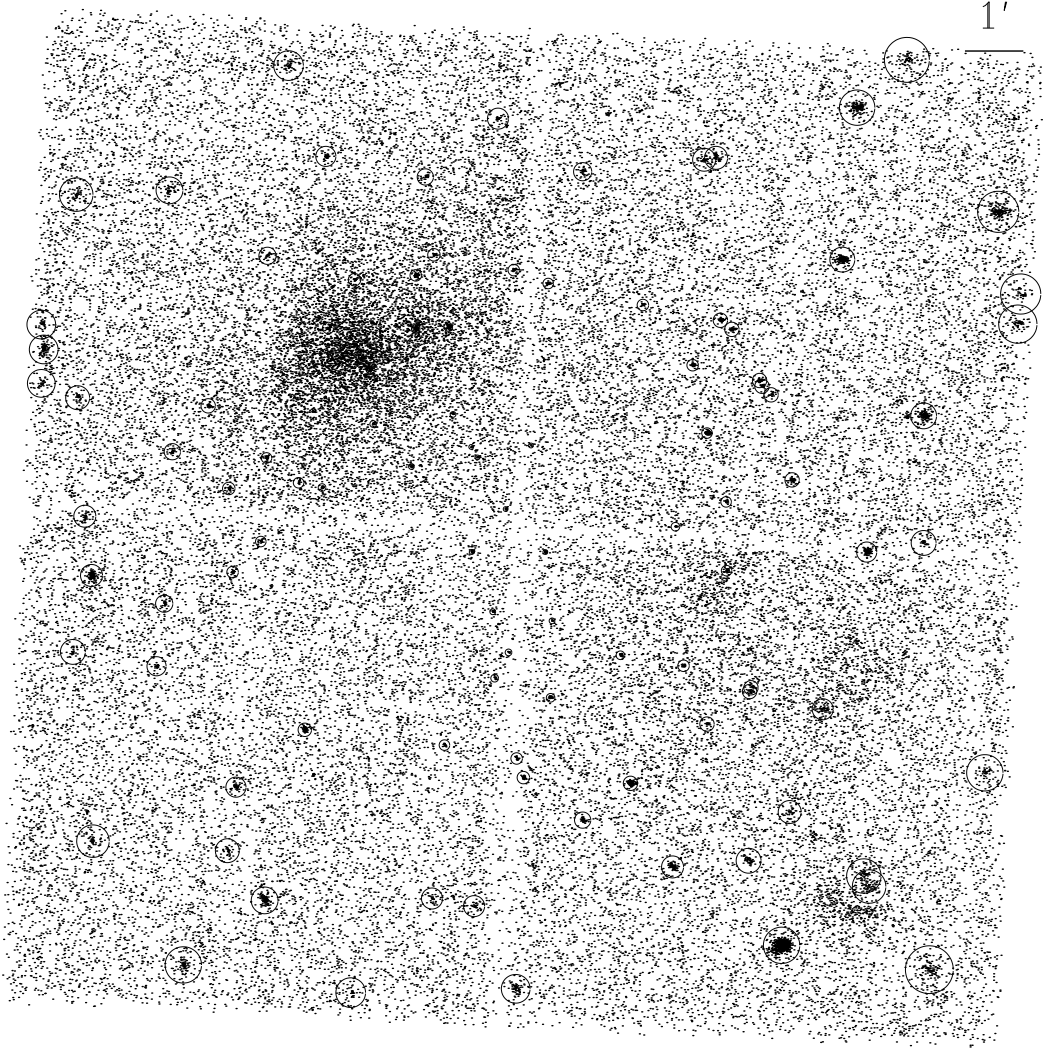,height=6.truein,angle=90.0,clip=}
}
\caption{\protect\footnotesize 
ACIS-I count image of Abell 2125 in the 0.5-8 keV band.
Detected X-ray sources are marked with circles of
twice the $\sim 90\%$  energy-encircled radii of the PSF
(Jerius et al. 2000), estimated in the same 
band for a power law spectrum with photon index equal to 1.7
and $N_H \sim 3 \times 10^{20} {\rm~cm^{-2}}$.}
\label{fig1}
\end{figure}

\begin{figure}[htb!]
\centerline{
\psfig{figure=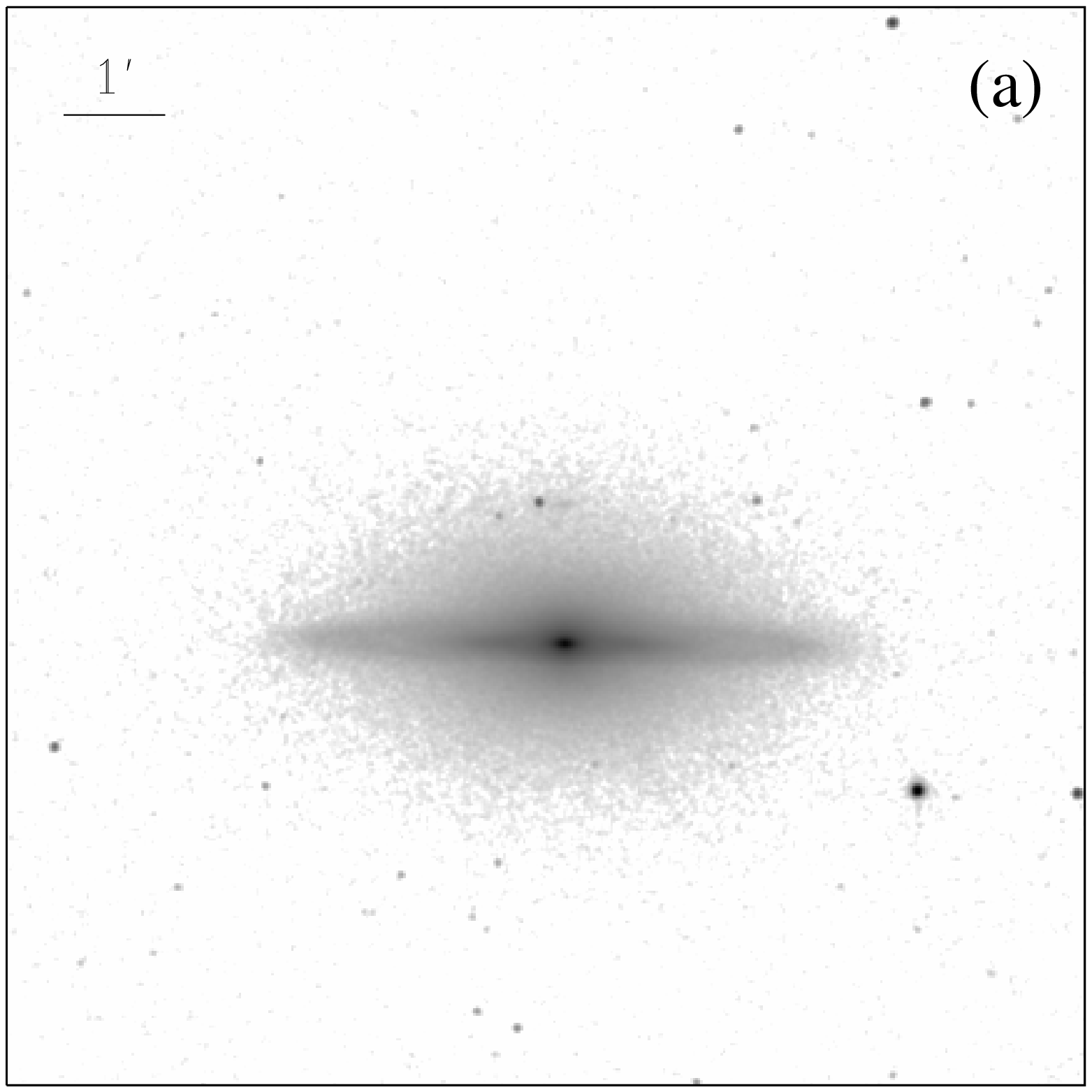,height=3.2truein,angle=90.0,clip=}
\psfig{figure=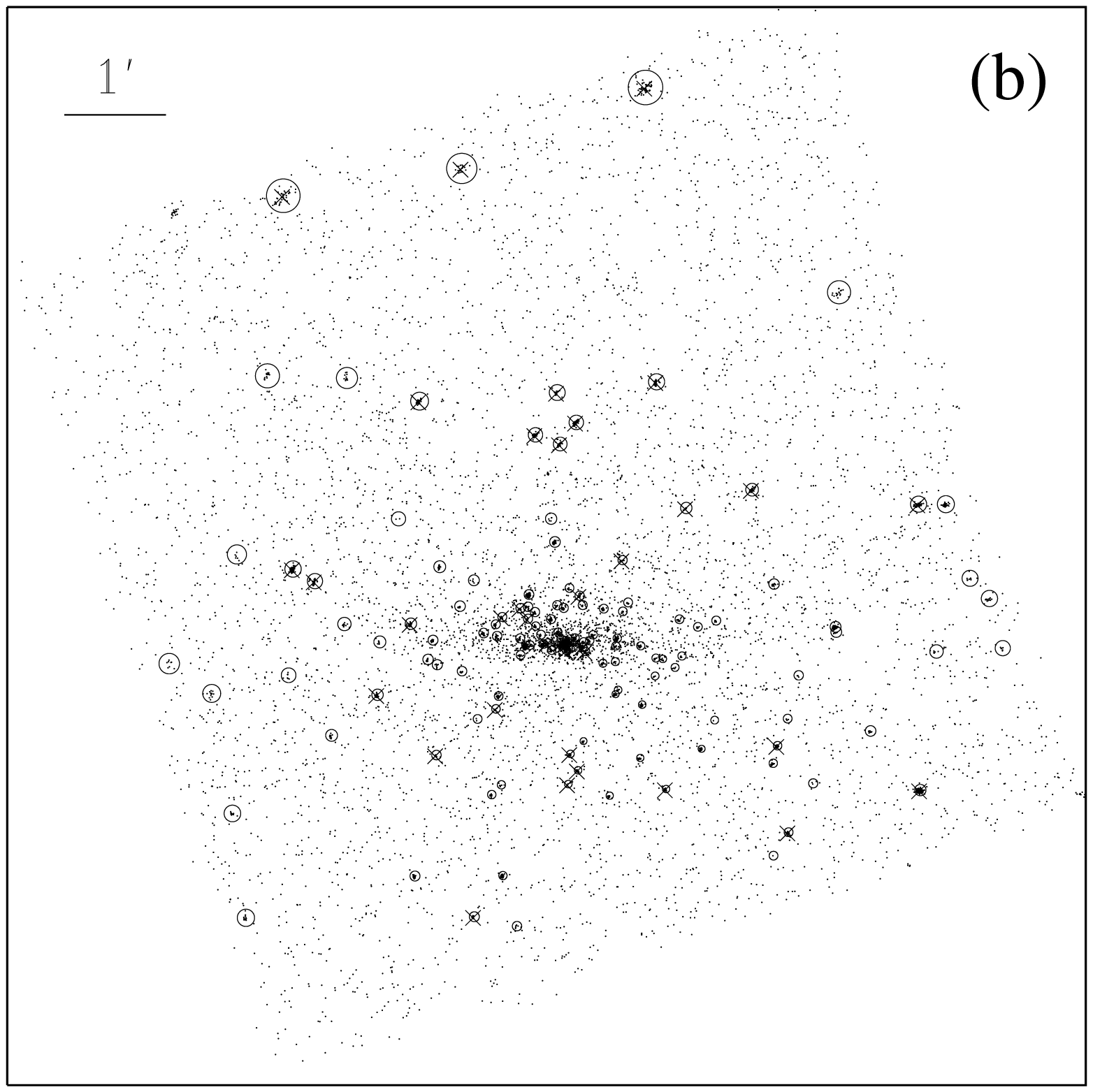,height=3.2truein,angle=90.0,clip=}
}
\caption{\protect\footnotesize  (a) {\sl 2MASS} K-band
image and  (b) ACIS-S count image of NGC4594
in the 0.3-7 keV band. 
Detected X-ray sources are marked with circles of
twice the $90\%$ EER. 
Sources identified as
globular clusters, stars, and central nucleus (all marked with $\times$) are
not included in the number-flux relation analysis (\S 4).} 
\label{fig1}
\end{figure}

Various source detection and analysis schemes (wavelet, sliding-box, 
and maximum likelihood centroid fitting) have been used in previous studies
(e.g., Freeman et al. 2002; Harnden et al. 1984; Cruddace et al. 1988). 
Applications of these schemes to {\sl Chandra} imaging 
data have been described in Wang et al. (2003; 2004).
To detect a source is to find a significant count number deviation
from the expected statistical background fluctuation: 
the total count number $n_c$ within a certain detection aperture is
compared with the expected contribution from the background $n_b$ (e.g., 
Fig. 3a), accounting for the statistical fluctuation. 
$n_b$ is typically estimated from the diffuse X-ray intensity in the region 
surrounding the detection aperture. Specifically, we first remove source
candidates identified in the wavelet detection with a reduced  
threshold ($S/N \sim 2$) and then smooth the remaining image with a Gaussian kernel 
adjusted adaptively to achieve a spatially uniform count-to-noise 
ratio of $\sim 10$. The exposure correction for the intensity calculation
accounts for such effects as the effective area variation, bad pixel/column 
removal, and CCD gaps. The resultant background intensity image is not
sensitive to the exact source-removal threshold and smoothing method.
The image, multiplied
by the exposure map, gives the background count image (e.g., Fig. 3a), from
which $n_b$ can be obtained within the corresponding detection aperture. 
The selected intensity-to-noise 
ratio in the smoothing is large enough so that 
the uncertainty in the background intensity estimate is small 
and may thus be neglected. The application
of such a background map in a source search is sometimes called 
``the map detect algorithm'', which is well-defined for a statistical analysis
of the source detection.

The significance of a count deviation above the background contribution $n_b$
can be characterized by the false detection probability $P$ defined as
\begin{equation} 
P =1- \sum_{n=0}^{n_c-1} {n_b^n \over n!} e^{-n_b}.
\end{equation} 
If $P < P_{th}$, a preset threshold, one may declare a positive source 
detection. This single threshold is normally sufficient, except for the 
situation in which $n_b$ is extremely small (e.g., in regions partially covered
by CCD gaps due to the telescope dithering; Fig. 1). Then an
additional condition needs to be imposed for a meaningful
detection (see further discussion in \S 3). 
The probability $P$, however, is for a single comparison only. 
In a typical blind search for sources, an entire X-ray image is scanned, 
invoking many statistically independent comparisons. Therefore,
the false detection probability in such a search is much greater than 
$P$ and depends on both the detection aperture and the size of the image as 
well as the choice of the threshold $P_{th}$. For example, searching in 
count images simulated by adding Poisson noises in the
background image shown in Fig. 3a yields an average of $\sim 0.05$ and 
0.6 fake detections for $P_{th} =10^{-6}$ and $10^{-5}$.

\begin{figure}[htb!]
\centerline{
\psfig{figure=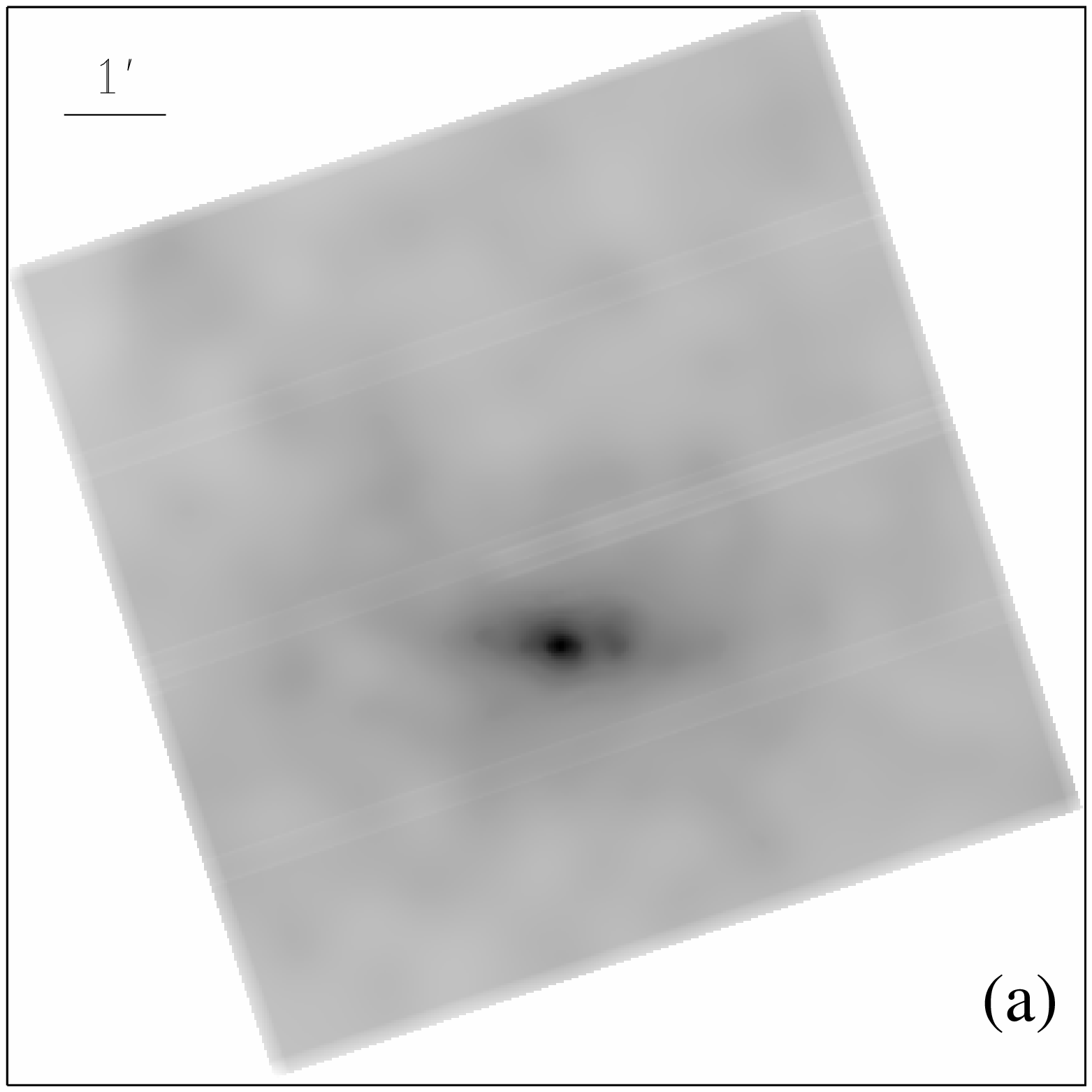,height=3.2truein,angle=90.0,clip=}
\psfig{figure=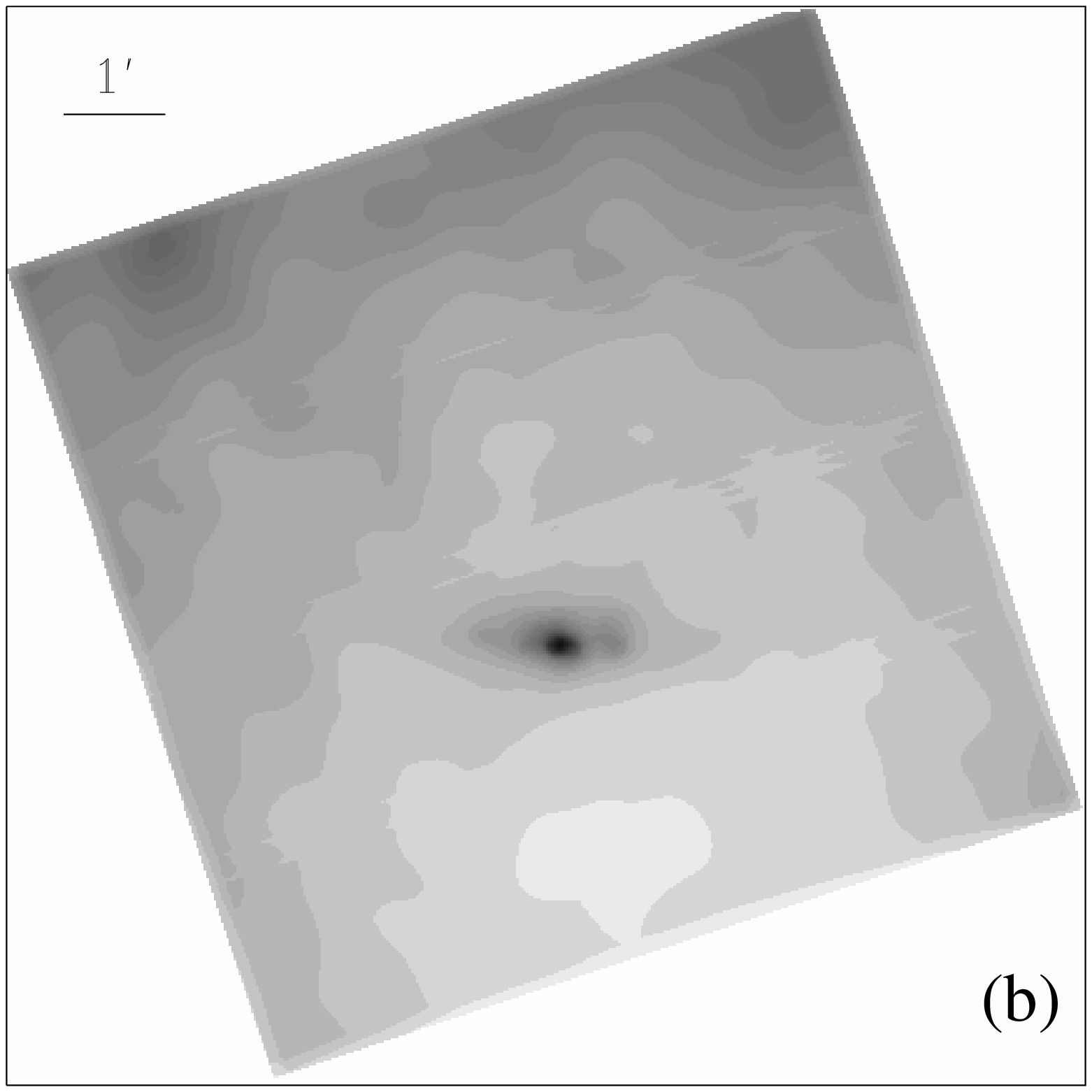,height=3.2truein,angle=90.0,clip=}
}
\caption{\protect\footnotesize (a) Smoothed background map of the 
count image as presented in Fig. 2b. The gray-scale is in the range 
from 0.01 (white) to  9.0 (black).
 (b) Source detection threshold map for 
$P_{th} =10^{-6}$, showing the minimum required number of net source counts 
(in the range of 4 -- 40)
 within the position-dependent source-detection aperture and above 
the background as shown in (a). }
\label{fig2}
\end{figure}

\begin{figure}[htb!]
\centerline{
\psfig{figure=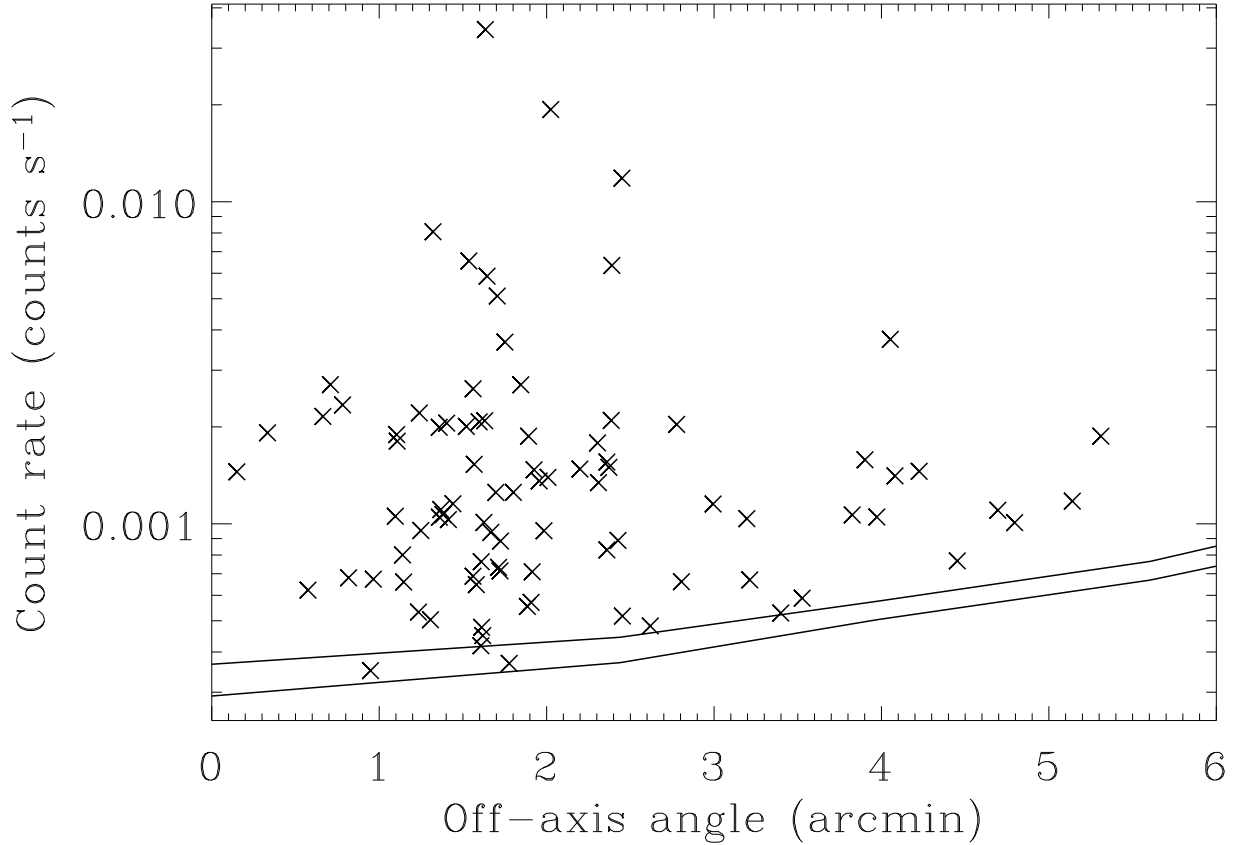,height=5.truein,angle=0,clip=}
}
\caption{\protect\footnotesize
Count rates of the sources detected in the 0.3-7 keV band versus their 
off-axis angles in the ACIS-S observation of NGC 4594. 
The curves illustrate the detection thresholds ($S_{min,k}$):
The upper curve is calculated via an azimuthal average, whereas
the lower curve is obtained by choosing the lowest value 
in each concentric annulus around the aiming point of the observation. A
similar plot for the Abell 2125 field is presented in Wang et al. (2004),
}
\label{fig4}
\end{figure}

For a detected source, one may estimate its count rate,
$S =  n_s/(\eta t)$, where $n_s = n_c -n_b$,  $t$,  and $\eta$
are respectively the net number of source counts, the effective exposure 
time, and the corresponding energy-encircled fraction of the 
detection aperture. The count rate can be easily converted 
into an energy flux if an X-ray spectrum is assumed. Thus the terms,
count rate and energy flux, are mostly interchangeable in the present work.
The count rate threshold is $S_{min,k} = 
n_{s,min,k}/(\eta t_k)$, where $n_{s,min}$ is the minimum 
number of source counts required for a positive detection. 
$S_{min,k}$ (or $n_{s,min}$; e.g., Fig. 3b) generally decreases with 
increasing off-axis angle, chiefly because of increasing PSF  size  
(e.g., Figs. 1 and 2b; Jerius et al. 2000) and because of the decrease 
in the effective exposure (due to telescope vignetting).
In the ACIS-S observation of NGC 4595 (Figs. 2 and 3), 
for example, the detection sensitivity is the highest at the aiming point 
(close to the axis of the telescope), which is about 
2$^\prime$ southwest of the galaxy's center (Fig. 3b). The sensitivity is 
substantially lower in the central region of the galaxy (Fig. 3b), 
because of the high diffuse X-ray background (Fig. 3a).
The sensitivity is also low at large off-axis angles primarily 
due to relatively large PSF sizes, especially near 
the northeastern CCD boundaries (Fig. 3b). 
Fig. 4 presents the detected source count rates, compared with
our calculated source detection thresholds.

The count rate, or flux, of a detected source represents only 
a single realization of the intrinsic source flux in the observation 
with limited counting statistics 
and is therefore subject to the Eddington bias.

\section{Maximum Likelihood Correction for the Eddington Bias}

Here we follow the approach described by Hogg \& Turner (1998) but
assume the Poissonian counting statistics. The likelihood 
that a source with an intrinsic count rate $S$ 
is observed to have a count rate $S_o$ is (see also Schmitt \& Maccacaro 1986)

\begin{equation} 
p(S_o|S) = {(n_s + n_b)^{n_c} \over n_c!} e^{-(n_s + n_b)}.
\end{equation}

\begin{figure}[htb!]
\centerline{
\psfig{figure=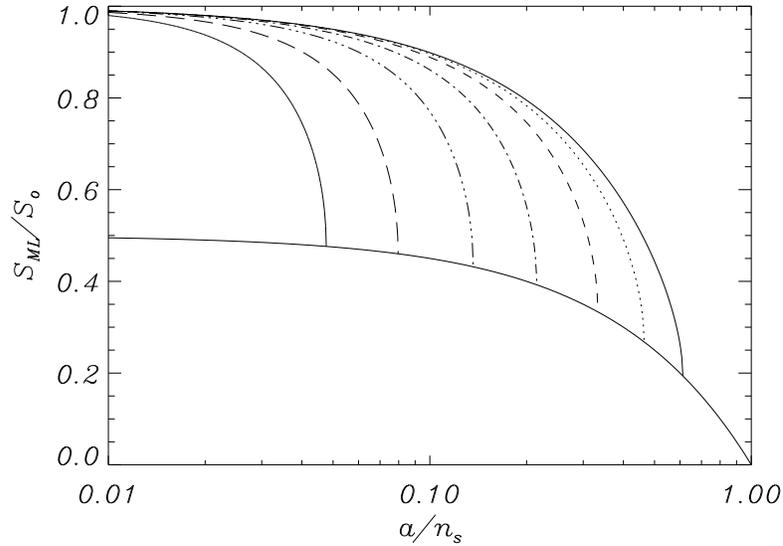,height=3.2truein,angle=0.0,clip=}
}
\caption{\protect\footnotesize
The dependence of the ML flux to the observed flux ratio 
on  $\beta = {\alpha \over n_s}$ 
and $\gamma = {\alpha \over n_b}$. The minimum $S_{ML}/S_o$
as a function of $\beta$ is represented by the thick curve, whereas
other curves from the left to 
the right are for $\gamma = $ 0.01, 0.03, 0.1, 0.3, 1, 3, and 10.
}
\label{fig5}
\end{figure}

The likelihood $p(S|S_o)$ for a true count rate $S$, given that this 
source is observed to have a count rate
$S_o$, is related to $p(S_o|S)$ by Bayes's theorem:
\begin{equation} 
p(S|S_o) \propto  p(S_o|S)f(S), 
\end{equation}
where $f(S)$ is 
the underlying intrinsic number-flux relation in the count rate interval from
$S$ to $S+dS$. 
If the count rate measurement were unbiased, the peak in the likelihood
function $p(S|S_o)$ would be at $S/S_o =1$. Because of the Eddington bias, 
however, $S_o$ is statistically greater than $S$. Over a relatively small 
range of $S$, considered here for the bias correction, it is generally a good 
approximation to assume 
$f(S) \propto S^{-\alpha}$, where the exponent $\alpha$ is a constant 
and is typically around 5/2 (the so-called Euclidean slope).
We take the derivative of $p(S|S_o)$ to find its ML peak position 
$S_{ML}$, which leads to 
\begin{equation} 
{S_{ML} \over S_o} = {1 \over 2}(1-\beta)\Bigl\{1+\Bigl[1-{4 \beta^2 
\over \gamma(1-\beta)^2}\Bigr]^{1/2}\Bigr\},
\end{equation}
where $\beta = {\alpha \over n_s}$ and $\gamma = {\alpha \over n_b}$.
Fig. 5 illustrates the dependence of the ratio ${S_{ML} \over S_o}$ on these two parameters; e.g., for a fixed  $\gamma $ the ratio decreases with 
increasing $\beta$. Evidently, the existence of the ML solution requires 
\begin{equation} 
\beta \leq \gamma^{1/2}/(2+\gamma^{1/2})
\end{equation}
or 
\begin{equation} 
n_s \geq \alpha+(4n_b\alpha)^{1/2}. 
\end{equation}
If $\alpha = 5/2$, 
\begin{equation} 
n_s \geq 2.5+(10n_b)^{1/2}. 
\end{equation}
For example, $n_s$ needs to be $\gtrsim 6$, 4, or 3  for $n_b \sim 1$, 0.1,
or  0.01 (corresponding
to log$(P) = -3.2$, $-5.4$, or $-6.8$; see Eq. 1). 
We force this condition to be satisfied for a positive source 
detection, in addition to $P < P_{th}$ (\S 2). Otherwise, a
detection with $n_s=1$ might be claimed as a source if $n_b$ is small enough
(see Eq. 1), which could occur especially at edges of CCDs.
The largest ML correction, or the smallest ratio 
${S_{ML} \over S_o}$, is (see also Fig. 5)
\begin{equation} 
{S_{ML} \over S_o} = {1 \over 2}(1-\beta).
\end{equation}

If $\beta \ll 1$ (i.e., $n_{s} \gg \alpha$), the solution (4) then becomes 

\begin{equation} 
{S_{ML} \over S_o} = {1 \over 2}\Bigl\{1+\Bigl[1-{4 \alpha 
\over (n_s/n_b^{1/2})^2}\Bigr]^{1/2}\Bigr\}.
\end{equation}
This solution has the same appearance as the one derived by Hogg 
\& Turner (1998) based on an assumed Gaussian error distribution. But
$n_s/n_b^{1/2}$ here is the signal-to-background noise ratio, instead of 
the signal-to-noise ratio $S_o/\sigma$, where $\sigma$ is the Gaussian
dispersion of $S_o$ and is assumed to be a constant (Hogg \& Turner 1998).
In X-ray astronomy, at least, the uncertainty in $S_o$ depends on $S_o$
itself, even when the Poissonian error distribution asymptotically approaches 
the Gaussian distribution for large $n_s$ and $n_b$. Furthermore,
the signal-to-background noise ratio  is  always 
substantially greater than the signal-to-noise ratio $n_s/(n_s+n_b)^{1/2}$
for the same $n_s$. Therefore, adopting the Poissonian
error distribution, which is appropriate for the X-ray counting
statistics, we can apply the ML correction to individual
sources detected with only a few counts, as long as Eq. 7 is satisfied.

\section{Analysis of the Number-Flux Relation}

The goal of this analysis is to constrain the intrinsic (model) 
distribution $f(S)$ over a broad range of $S$, 
based on an observed number-flux distribution.
Integrating Eq. (3) with a proper normalization included gives the expected 
number of sources in the $S_o$ to $S_o+dS_o$ interval
\begin{equation} 
N(S_o) = \int_0^\infty R(S_o,S)f(S)dS,
\end{equation}
where $R(S_o,S)$ is the probability that a source with an intrinsic 
count rate $S$ is expected to have a detected count rate in the 
interval. 
The above $N(S_o)$ can be
used directly in a ML analysis of the observed source
number-flux distribution to constrain model parameters in $f(S)$. 

However, it is often convenient
to convert the above equation into a discrete expression with $S$ and $S_o$
divided into channels $I$ and $J$ of small count rate widths, $\Delta_I
=S_I^{u}-S_I^{l}$ and $\Delta_J = S_J^u-S_J^l$, where $l$ and $u$ denote 
the lower and upper count rate boundaries of the channels. We
define the channels evenly on the logarithmic scale  of source count rate. The
overall spans of the channels should be large enough to cover the count rate
range involved. For example, we choose the lowest $I$ 
channel to be a factor of 10 less than the lowest flux threshold of
the source detection. By integrating over each channel $J$ on both sides of
Eq. 10, we get 

\begin{equation} 
N(J) = \sum_{I} R(J,I) f(I) \Delta_I,
\end{equation}
where 
\begin{equation} 
f(I) = \int_{S_I^l}^{S_I^u} f(S) dS/\Delta_I
\end{equation} 
and the redistribution matrix
\begin{equation} 
R(J,I) = {\int_{S_I^l}^{S_I^u} R(J,S) f(S) dS \over \int_{S_I^l}^{S_I^u} f(S)dS},
\end{equation}
in which $R(J,S) = \int_{S_J^l}^{S_J^u} R(S_o,S) dS_o$. In practice,
$\Delta_I$ can be chosen to be small enough so that 
\begin{equation} 
R(J,I) \approx  R(J,S_I),
\end{equation}
where $S_I$ may be defined as $S_I = (S_l S_u)^{1/2}$.

As discussed in \S 3, the 
redistribution matrix is basically the Poisson probability 
distribution, although an average is required over the field from which 
the sources are 
detected. This step is similar to the construction of a weighted instrument 
redistribution function for the spectral analysis of a diffuse X-ray 
feature. In the present case, the average is needed because 
the source detection sensitivity varies across the field. The 
matrix defined at each pixel $k$ can be expressed as 
\begin{equation} 
R_k(J,S_I) = \left\{ \begin{array}{ll} 0 & \textrm{if $S_J^u \leq S_{min,k}$}\\
         Q_k(<S_J^{u},S_I)-Q_k(<S_J^{l},S_I) & 
         \textrm{if $S_J^l \geq S_{min,k}$}\\
         Q_k(<S_J^{u},S_I)-Q_k(<S_{min,k},S_I) & 
         \textrm{if $S_J^l < S_{min,k} < S_J^u$}
\end{array} \right.
\end{equation} 
where $Q_k$ is the accumulated Poisson probability (see also Eq. 2). For example,
\begin{equation} 
Q_k(<S_J^u,S_I)=\sum_{n=0}^{n_{max}} 
{(\eta t_k S_I+n_{b,k})^n \over n!} e^{-(\eta t_k S_I+n_{b,k})},
\end{equation} 
where  $n_{max}$ is the largest integer that is smaller than 
$\eta t_k S_J^u+n_{b,k}$. When the expected value of $\eta t_k S_I+n_{b,k}$ is large, 
$Q_k$ can be replaced by the standard error function of a Gaussian
distribution. The calculation
does not need to be done for all $J$ channels,
as long as the range included covers essentially all the redistribution
probability; a 4$\sigma$ range around each $S_I$ is calculated in our applications.
XSPEC, for example, allows for a redistribution matrix 
with a variable $J$ channel dimension. The redistribution matrix averaged
over all pixels is then 
\begin{equation} 
R(J,I) = \sum_{k=1}^{N_p} R_k(J,I)\xi_k,
\end{equation}
and
\begin{equation} 
\sum_{k=1}^{N_p} \xi_k=1,
\end{equation} 
where $\xi_k$ 
is the intrinsic probability for a source to appear at the $k$th pixel
and 
may be inferred from their empirical spatial distribution, from 
a model, and/or from an image in another wavelength
band (e.g., optical). For example, to estimate the total number of background 
sources in a field $N_{s,b}$, we can 
assume a uniform probability distribution (i.e., $\xi_k = 1/N_p$). 
On the other hand, we may assume that a source population of
galaxy follows its near-infrared light (e.g., Fig. 2a) 
with an intensity $I_k$ at pixel $k$. Under these assumptions, we have
\begin{equation} 
 \xi_k = {N_{s,b} \over N_s N_p} + (1- {N_{s,b} \over N_s}){I_k \over \sum_{k=1}^{N_p} I_k},
\end{equation} 
where $N_s$ is the total number of detected sources, background plus galactic.

The above defined $R(J,I)$ accounts for both the 
detection threshold variation and the flux redistribution of the sources.
Fig. 6 shows an example of the redistribution matrix. As expected, the
Eddington bias causes a systematic shift of the probability distribution 
peak (for a fixed $S$) to the right of $S_o=S$, apparent for 
$S \lesssim 10^{-3} {\rm~counts~s^{-1}}$. The dispersion in $S_o$ 
also increases with decreasing $S$. Effects due to the 
discreteness of the Poissonian 
distribution are apparent at low $S_o$ values. Fig. 7 illustrates these effects
for three representative $S$ values. Furthermore,
the sum of  $R(J,I)$ over $J$ channels gives the average 
detection probability
as a function of intrinsic source count rate (e.g., Fig. 8).
\begin{figure}[htb!]
\centerline{
\psfig{figure=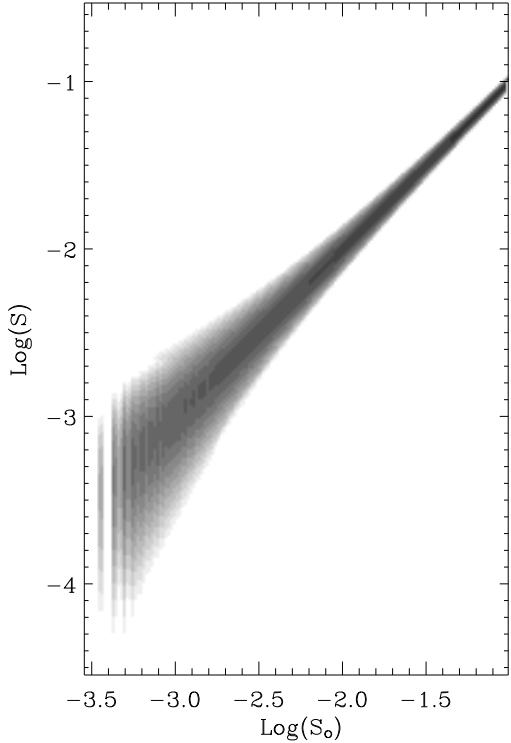,height=4.truein,angle=0,clip=}
}
\caption{\protect\footnotesize
Source count rate distribution matrix in the 0.3-7 keV band for the
NGC 4594 observation (see also Figs. 2-3; see also \S 5). The probability is scaled 
from $1 \times 10^{-5}$ (white) to 0.4 (black), logarithmically.
Both $S$ and $S_o$ are in units of ${\rm~counts~s^{-1}}$.
}
\label{fig6}
\end{figure}

\begin{figure}[htb!]
\centerline{
\psfig{figure=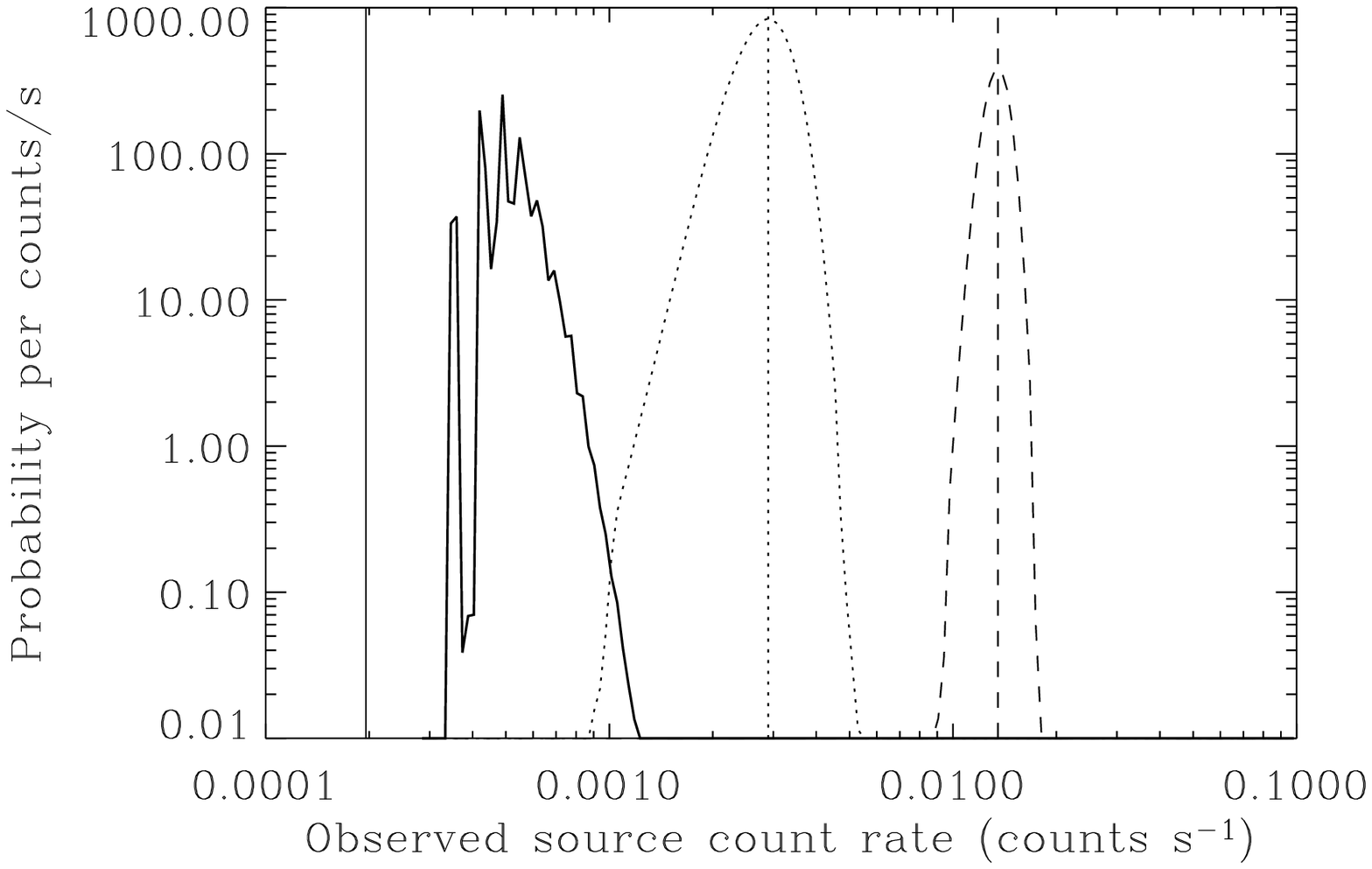,height=4.truein,angle=0.0,clip=}
}
\caption{\protect\footnotesize
Illustration of the average source flux redistribution probability in the 
NGC 4594 observation. The vertical straight lines mark the selected intrinsic 
source fluxes ($S$), whereas the curves represent 
the probability distribution of the corresponding flux ($S_o$). 
Notice that the solid line marks an $S$ value
that is below the lowest detection threshold of $\sim 3 \times 10^{-4} {\rm~counts~s^{-1}}$.
}
\label{fig7}
\end{figure}

$R(J,I)$, together with the count rate boundaries
of the $I$ and $J$ channels (all included in a standard fits file), can be
imported into XSPEC. The only other file required is the 
histogram of the detected sources in the $J$ channels. 
All other calculations, such as the binning of the model distribution, 
are performed automatically  within XSPEC, which also allows for using 
$\chi^2$ or Cash's C-statistic in the model-fitting and for 
estimating model parameter uncertainties.

The model distribution $f(S)$ can essentially be in any form, e.g., 
an additive and/or multiplicative combination of various components.
Different components may have different count rate-to-energy flux 
conversions, which depend on the line-of-sight
absorption/scattering as well as the assumed intrinsic source spectra. 
Allowing for this flexibility is the primary reason
for us to use count rates, 
instead of energy fluxes here. If a single conversion (i.e., a single
incident spectrum) is assumed for all sources,
expressing the procedure  described above in terms of energy fluxes is trivial.


\begin{figure}[htb!]
\centerline{
\psfig{figure=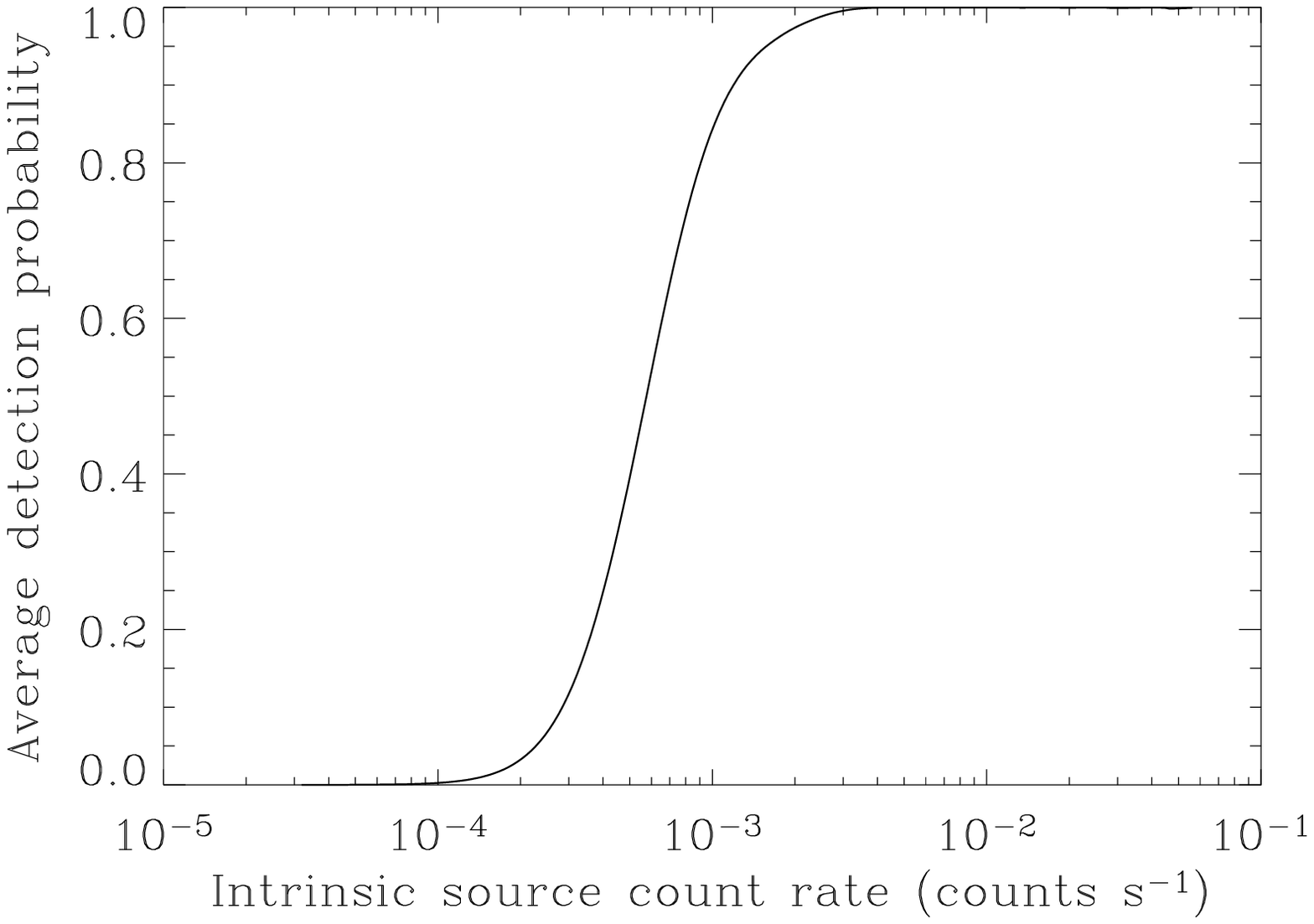,height=4.truein,angle=0.0,clip=}
}
\caption{\protect\footnotesize
Average detection probability of a source as a function of its intrinsic 
count rate in the  NGC 4594 observation.
}
\label{fig8}
\end{figure}
 
\section{Applications}

While the ML Eddington bias estimate for individual sources is straight 
forward, we concentrate on the application of the source flux redistribution
matrix. This procedure follows the standard data calibration, 
exposure map construction,
source detection, and background image generation (e.g., Wang et al.
2003, 2004; see also Figs. 1-3).  We adopt $P_{th} = 10^{-6}$.
From the ACIS-I observation of Abell 2125,
we detect 81, 48, and 93 sources in the 0.5-2, 2-8, and 0.5-8 keV bands, respectively.
From the ACIS-S observation of NGC 4594, 
80, 65, and 112 sources are detected 
in the 0.3-1.5, 1.5-7, and 0.3-7 keV bands, 
respectively. After removing duplicates (spatial coincidences)
in the detections, we find a total of 99 and 115
unique sources in the Abell 2125 and NGC 4594 fields (Figs. 1 and 2b).
We then apply the procedure described in
\S 4 to generate the required flux redistribution matrix
$R(J,I)$ for each source detection band (e.g., Fig. 6), which can then be used
to study the number-flux relations in the fields.

\subsection{ Contribution from Interlopers}

In a typical {\sl Chandra} imaging observation, a considerable fraction, 
if not most, of the detected sources are interlopers. 
In a high Galactic latitude field, interlopers are typically
background AGNs, plus a small number of foreground stars. The
number-flux relation of such interlopers has been characterized in 
various X-ray surveys, including the {\sl Chandra} ACIS-I
deep surveys (Moretti et al. 2003 and 
references therein). In almost all recent work, this characterization is 
done separately in the 0.5-2 keV and 2-10 keV bands. The 
source populations detected in these two bands only partially overlap:
Sources with steep X-ray spectra 
preferentially appear in the 0.5-2 keV band, whereas 
highly-absorbed ones appear in the 2-10 keV band. In general, {\sl Chandra}
is much more sensitive to sources in the former band than in
the latter band. Therefore, we focus on the accumulated number-flux relation 
in the 0.5-2 keV band, which can be modeled as:
\begin{equation} 
N(>S) = N_0 \Bigl[{(2 \times 10^{-15})^{\alpha_1} \over S^{\alpha_1}
+S_0^{\alpha_1-\alpha2} S^{\alpha_2}}\Bigr],
\end{equation}
where $N_0 = 6150 {\rm~sources~deg^{-2}}$, $\alpha_1 = 1.82$,  
$\alpha_2 = 0.60$, and $S_0 = 1.48 \times 10^{-14} {\rm~erg~s^{-1}~cm^{-2}}$
(Moretti et al. 2003). 
We have implemented this model in
XSPEC, which automatically computes $f(I)=[N(>S_I^l)-N(>S_I^u)] A$,
where $A$ in units of deg$^2$ is the FoV of the observation in
consideration.
By convolving  $f(I)$ with $R(J,I)$, we may  predict the
number-flux relation of the interlopers (Eq. 11).

\begin{figure}[htb!]
\centerline{
\psfig{figure=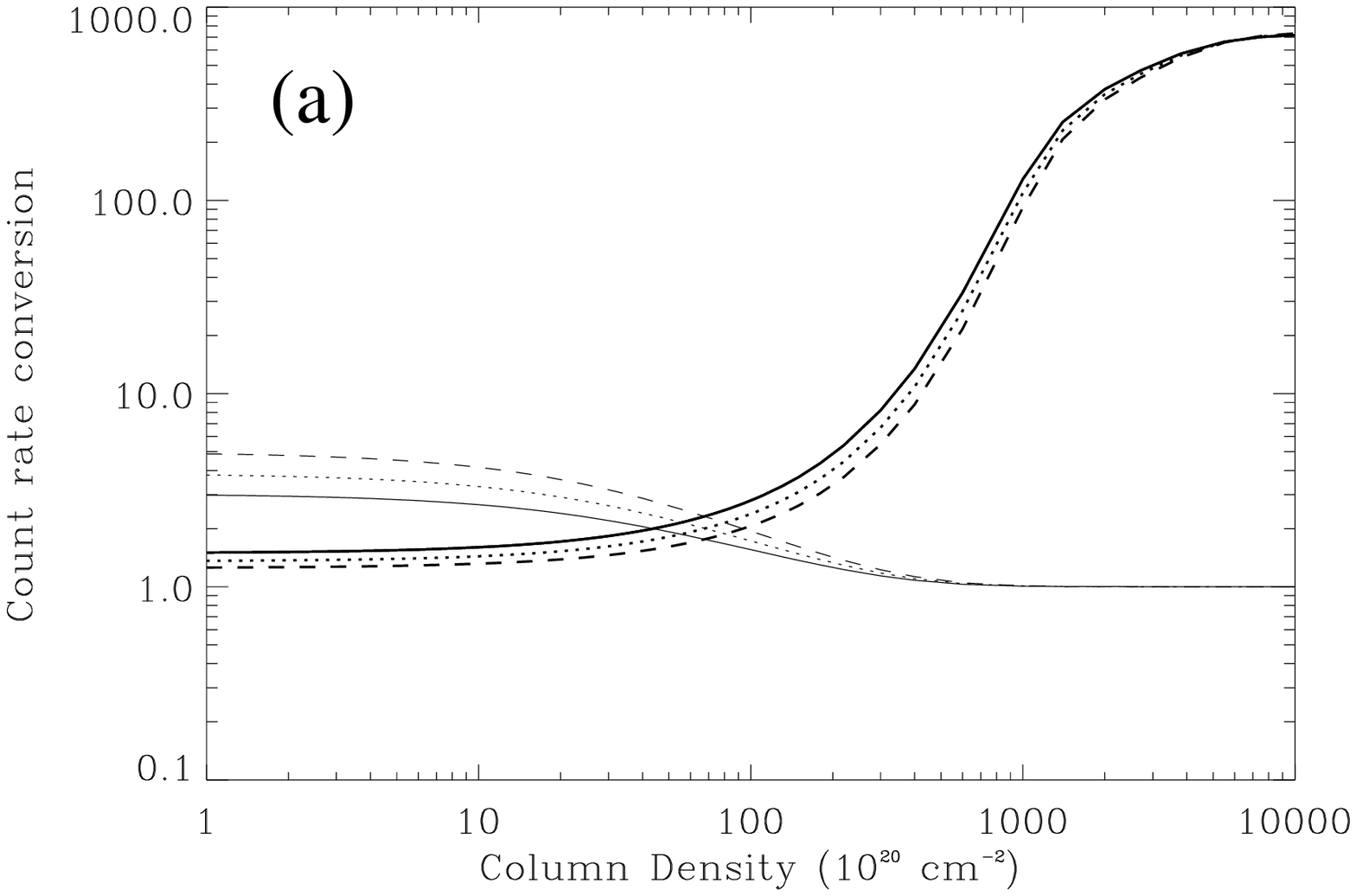,height=2.4truein,angle=0.0,clip=}
\psfig{figure=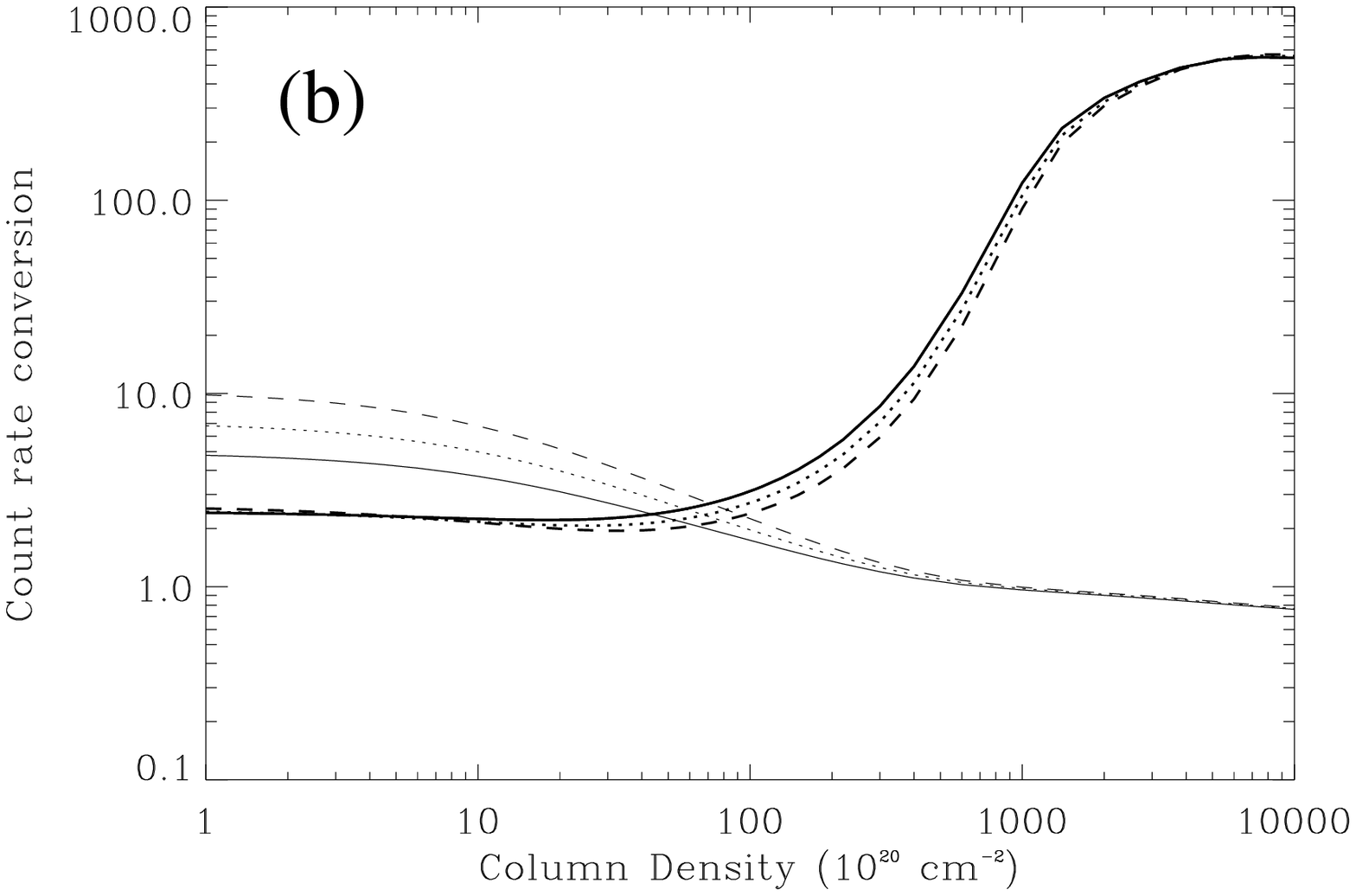,height=2.4truein,angle=0.0,clip=}
}
\caption{\protect\footnotesize The on-axis count rate ratio in different
bands versus X-ray-absorbing gas column density for a representative power-law models. (a) The ACIS-I ratios: 0.5-8 keV to 0.5-2 keV (thick curves) and
0.5-8 keV to 2-8 keV (thin curves). (b)
The ratios: ACIS-S 0.3-7 keV to ACIS-I 0.5-2 keV (thick curves) and
ACIS-S 0.3-7 keV to ACIS-I 2-8 keV (thin curves). The three styles of
the curves correspond to photon index choices: 1.4 (solid), 1.7 (dotted), and 
2 (dashed).} 
\label{fig9}
\end{figure}

However, caution must be exercised in using Eq. 20.
The energy flux $S$ specifically depends on 
the count rate-to-energy flux conversion that {\sl assumes} a power law 
with a photon index of 1.4 plus a foreground absorption of $1.6 \times 10^{20} 
{\rm~cm^{-2}}$. Indeed, the power law is a good description of the
overall spectral shape of the extragalactic X-ray background, which
is now believed to be a composite of mostly AGNs. But,
X-ray spectral properties do vary greatly from one source to another. 
Statistically, the average X-ray spectrum of AGNs tends to become flatter with
decreasing flux, apparently due to an increasing population of 
highly-absorbed AGNs. Unfortunately, such spectral variations are not yet well
quantified and thus cannot be incorporated easily into
the calculation of the count rate-to-energy 
flux conversion. Obviously, a deviation from the 
assumed source spectrum could cause errors.
To minimize this effect, we convert the energy 
flux in the model back into the count rate in the 0.5-2 keV band 
of an ACIS-I observation, using the same assumed power law, but with
the foreground absorption appropriate to a particular field. In the case
of the Abell 2125 observation, for example, we use 
$N_H = 3 \times 10^{20} {\rm~cm^{-2}}$ to derive the conversion as 
$4.5 \times 10^{-12}$ ${\rm~(erg~cm^{-2}~s^{-1}})/({\rm counts~s^{-1}})$,
which leads to a predicted number-flux relation of the interlopers, as 
is shown in Wang et al (2004). 

To construct the observed number-flux relation, we exclude the 10 
sources that have been found to be in positional coincidence
with Abell 2125 complex member galaxies (Wang et al. 2004).
These members are all detected in both 0.5-2 keV and 0.5-8 keV bands.
We construct the observed relation in the 0.5-2 keV band by grouping the 
remaining 71 sources
to have a minimum number of $\sim 8$ sources per count rate bin. 
A direct comparison between the observed and predicted relations
(not a fit) gives a C-statistic value of 6.9 for a total of 
9 bins. The probability 
to have a C-statistic less than this value is 38\%, according to 
Monte-Carlo simulations in XSPEC.
Therefore, the two relations are statistically consistent with 
each other.

A source number-flux relation comparison may also be made for detections 
in other bands. This comparison is particularly desirable for
the ACIS-I 0.5-8 keV band, in which the source detection is 
typically most sensitive. However, we are not aware of any 
number-flux relation constructed recently for interlopers in this band. 
Indirectly, one may still use Eq. 20  by converting the flux to
the count rate in the 0.5-8 keV band.  This approach implicitly uses
the ratio of the 0.5-8 keV to 0.5-2 keV count rates,
which depends on an assumed source spectrum (Fig. 9a). 
For a typical AGN spectrum with a power-law photon index between 1.4-2
and an absorption smaller than $\sim 10^{22} {\rm~cm^{-2}}$, 
a good approximation for the ratio is 1.3.
With this approximation, the comparison between
the observed and expected number-flux relations of the Abell 2125 sources
in the 0.5-8 keV band gives a C-statistic value of 18.4 for 10 bins,
which may be rejected with a confidence of $\sim 94\%$.
The predicted number of sources is 75, compared to 83 actual detections
(the complex members are excluded). This difference is partly
due to the expected presence of relatively hard X-ray sources, which are 
not detected in the 0.5-2 keV band and are thus not included in the
number-flux relation extrapolated from the same band. Unfortunately, 
the poor number statistics of such sources detected in the field 
prevents us from placing a tight constraint on their population. 
In addition, one also expects an intrinsic cosmic variance 
in the source number density from one field to another,
typically $\gtrsim 6\%$, depending on the FoV, energy band, and source
detection limit of observations (see Yang et al. 2003 and references therein).
Therefore, even in a relatively deep exposure such as
the Abell 2125 observation,  the number of interlopers
cannot be predicted accurately. 

The relative uncertainty in the number of interlopers
becomes even greater for an ACIS-S observation, because of
both its small FoV (at least for the BI CCD \#7 chip alone) and its 
different energy response from the FI CCDs of the ACIS-I. There is not
yet a number-flux relation constructed for ACIS-S detected interlopers. 
Nevertheless, an approximate 
estimate is often required. One may use an
approach similar to the one  described above for the ACIS-I 0.5-8 keV band.
By adopting a ratio of the ACIS-S 0.3-7 keV to ACIS-I 0.5-2 keV 
count rates as $\sim 2.2$ (Fig. 9b), for example, we find that the expected number
of interlopers is $N_{s,b} \sim 16$ in the NGC 4594 observation, compared with 112 sources 
detected in the 0.3-7 keV band. Although the relative uncertainty 
in the expected number of interlopers is statistically large, 
the bulk of the detected sources are clearly associated with the galaxy.

\subsection{Number-Flux Relation of X-ray Sources in NGC 4594}

To facilitate a comparison with the analysis by Di Stefano et al. (2003), we 
also exclude sources that have been identified 
as foreground stars as well as the nucleus and globular clusters of 
NGC 4594. Fig. 10 presents the number-flux relation for the remaining 
$N_s = 90$ sources detected in the 0.3-7 keV band. 
We calculate the re-distribution matrix, by using Eq. 19 with
$N_{s,b} = 16$ as estimated above and the 2MASS K-band image (Fig. 2a)
as the galactic source
probability distribution. 
Foreground stars in the image 
are removed approximately with a 9$\times$9 pixel median filter 
(pixel size = 1\farcs5). The predicted interloper contribution 
is small and is thus fixed in the subsequent 
number-flux analysis. First, we model the galactic 
component, using a single power law and get the best fit as 
\begin{equation} 
{\rm Log}[N(S)] = -1.6_{-0.39}^{+0.38}-2.1_{-0.12}^{+0.14}{\rm Log}(S)
\end{equation} 
(all error bars are at the $1\sigma$ confidence level);
the C-statistic value 17.0 for 9 degrees of freedom can only be 
rejected statistically at a confidence $\sim 92\%$. Considering
all the additional uncertainties that are not included in the analysis
(e.g., interloper subtraction errors),
we consider the single power law fit is still reasonably acceptable. 

Next, we fit the observed number-flux relation with a broken power law
(see also Di Stefano et al. 2003),
\begin{equation} 
\left({dN \over dS}\right)=N_b\left({S \over S_b}\right)^{-\Gamma},
\end{equation} 
where $\Gamma = \Gamma_1$ for $S \leq S_b$ and 
$\Gamma = \Gamma_2$ for $S > S_b$.
 With an addition of only two more model parameters than in the single power law,
this model fit gives a C-statistic value of 8.87, which is satisfactory
(Fig. 10). The best-fit parameters are 
$N_b=1.7 \times 10^4 {\rm~sources/(counts~s^{-1}})$, $S_b = 2.1_{-0.7}^{+1.1}\times 10^{-3}$, 
$\Gamma_1 = 1.4_{-0.3}^{+0.3}$, $\Gamma_2 = 4.6_{-1.2}^{+2.7}$. Fig. 11 presents
the confidence contours, $\Gamma_1$ vs.  $\Gamma_2$.

\begin{figure}[htb!]
\centerline{
\psfig{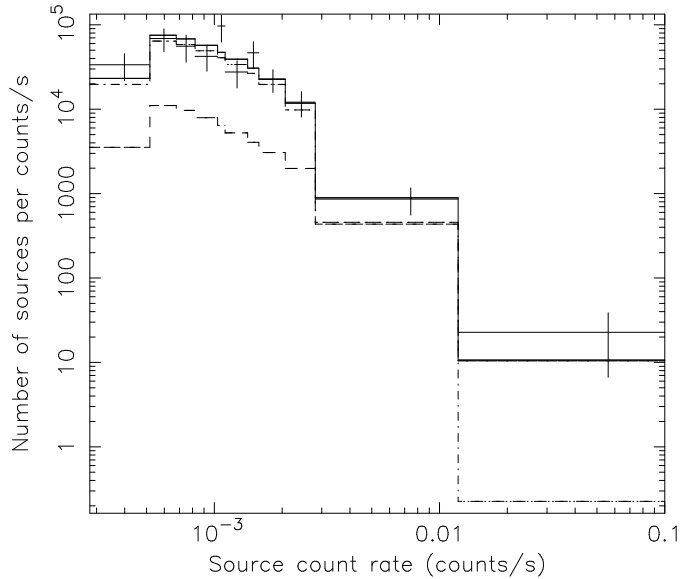}
}
\caption{\protect\footnotesize
Observed differential number-flux relation of the sources
detected in the 0.3-7 keV band, compared with the best-fit 
broken power-law model (dash-dotted histogram) plus the
expected interloper contribution (dashed) in the field of NGC 4594. 
The combination of the two model components is represented by the solid 
histogram. 
}
\label{fig10}
\end{figure}

\begin{figure}[htb!]
\centerline{
\psfig{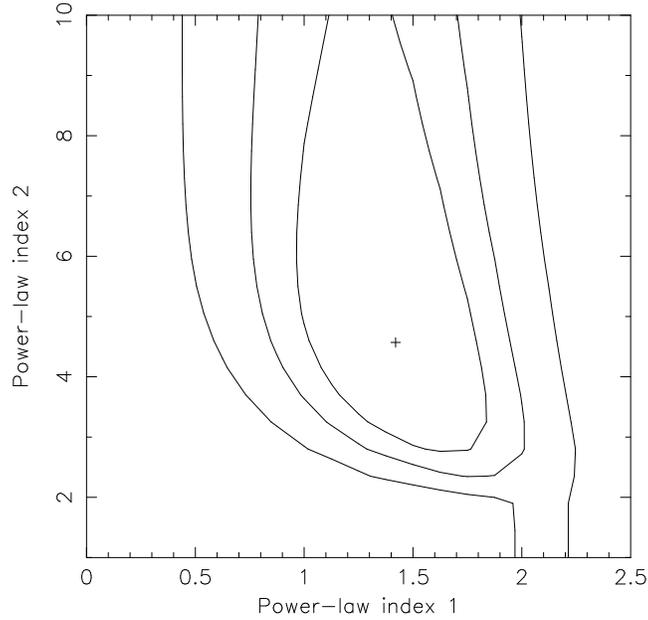}
}
\caption{\protect\footnotesize
The 68\%, 90\%, and 99\% confidence contours of the 
broken power law indexes from the best fit shown in Fig. 10.
}
\label{fig11}
\end{figure} 

In comparison, the analysis by Di Stefano et al. (2003) suggests
$\Gamma_1 = 0.74_{-1.04}^{+0.84}$, $\Gamma_2 = 2.58_{-0.28}^{+0.35}$,
and a luminosity
break at $9_{-1.6}^{+2.0} \times 10^{37} {\rm~erg~s^{-1}}$, corresponding 
to a count rate break $S_b \sim 1.8_{-0.32}^{+0.40} 
\times 10^{-3} {\rm~counts~s^{-1}}$, assuming 
a power law spectrum of photon index 2 and an absorption of $N_H =
5 \times 10^{20} {\rm~cm^{-2}}$  (Di Stefano et al. 2003). 
While these parameters are marginally consistent with those obtained 
from our analysis, their constraints on $S_b$ seem to be
considerably tighter than ours.

This difference is probably largely due to the simplification
made in the analysis by Di Stefano et al. (2003). Their analysis ignored
both the background source contribution and the Eddington bias and
assumed a single source-detection threshold
of $4.5 \times 10^{37} {\rm~erg~s^{-1}}$, or a count rate of $ 9 \times
10^{-4} {\rm~counts~s^{-1}}$, inferred from 
the peak at $S/N \sim 3.2$ in a number-$S/N$ histogram of the detected
sources. However, Fig. 8 shows that the average detection probability is 
still only $\sim 80\%$ at this count rate. A detection 
probability 
$\gtrsim 95\%$, for example, requires a source count rate $\gtrsim 1.4 \times
10^{-3} {\rm~counts~s^{-1}}$. Clearly, part of the turn-over in the
observed number-flux relation at $S_b$ (Fig. 10) is due to the 
flux-dependent source detection threshold, which should be accounted
for in the modeling. Setting a threshold
at a higher count rate should have reduced this flux dependence. But 
more than half of the sources would then be excluded in the analysis.

Of course, the flux-dependence of the threshold is a result of the
detection sensitivity variation across the field.
The variation is even greater for an ACIS-I observation,
because of the larger off-axis FoV. Therefore, a proper correction for the 
flux-dependent threshold is 
essential to the analysis of the source number-flux relation, especially to
the full utilization of detected sources.

\section{Summary}

We have examined the detection threshold and 
the Eddington bias in flux measurements for sources identified 
with limited Poisson counting statistics.
Our main results are as follows:
\begin{itemize}

\item We have derived a ML correction for the
flux measurement bias, based on a Poissonian
error distribution. 
This correction can be applied to individual 
sources detected with very limited counting statistics.
The condition
for the presence of the ML correction (Eq. 7) should be required
for a source detection, in addition to the probability threshold.

\item The source detection threshold varies strongly with 
position, especially with the off-axis angle. A flux-dependent 
correction for this threshold  variation is essential to 
the statistical analysis of detected sources. 

\item We have developed a simple procedure to calculate the flux 
redistribution matrix, based on the exposure and diffuse 
background maps obtained directly from an observation. The matrix
inherently accounts for both the source detection threshold variation
and the Eddington bias and can be imported into a software 
package such as XSPEC so that the number-flux relation can be analyzed 
as if it were an X-ray spectrum.

\item We have applied the procedure to a deep ACIS-I observation
of the Abell 2125 complex. With 10 identified complex members
excluded, the 
remaining X-ray sources show a number-flux relation that is 
consistent with the expected interlopers of the
field.

\item We have also analyzed the number-flux relation of
X-ray sources detected in an ACIS-S observation
of NGC 4594. While the contribution from the expected 
interlopers is shown to be small, we have characterized the 
 number-flux relation of the galactic sources, using either a power law or
a broken power law. The latter model gives a considerably better 
characterization than the former. 
\end{itemize}

The techniques described in this paper can be easily implemented; an 
IDL procedure for generating the flux-redistribution matrix, for example,
can be obtained from the author.

\acknowledgements 
David Smith is acknowledged for help in writing programs that convert 
the calculated redistribution matrix and the background source 
number-flux relation into the XSPEC-required formats. This work was funded by 
NASA under the grants NAG5-10783, NAG5-9429, NAG5--8999, and GO1-2126.

\end{document}